\documentclass[a4paper,12pt]{article}
\usepackage{graphicx}
\usepackage{amssymb}
\usepackage{amsmath}
\begin{document}



\title{In-situ gamma spectrometry measurements of time-dependent Xenon-135 inventory in the TRIGA Mark II reactor Vienna}


\author{Julia Riede, Helmuth Boeck\\{\small TU Wien, Atominstitut, A-1020 Wien, Stadionallee 2}}
{\small \date{June 16th, 2013}}
\maketitle

\begin{abstract}
In this work, it has been shown that the time dependent $^{135}$Xe inventory in the TRIGA Mark II reactor in Vienna, Austria can be measured via gamma spectrometry even in the presence of strong background radiation. It is focussing on the measurement of (but not limited to) the nuclide $^{135}$Xe.

The time dependent $^{135}$Xe inventory of the TRIGA Mark II reactor Vienna has been measured using a temporary beam line between one fuel element of the core placed onto the thermal column after shutdown and a detector system located just above the water surface of the reactor tank. For the duration of one week, multiple gamma ray spectra were recorded automatically, starting each afternoon after reactor shutdown until the next morning. One measurement series has been recorded over the weekend. 

The  $^{135}$Xe peaks were extracted from a total of 1227 recorded spectra using an automated peak search algorithm and analysed for their time-dependent properties. Although the background gamma radiation present in the core after shutdown was large especially in the lower energy range, the  $^{135}$Xe peak located at 249.8 keV could be extracted from the most spectra where present and could be compared to theoretical calculations.\\
\emph{Keywords: }reactor, core, Xenon, gamma spectrometry, reactor poison
\end{abstract}
% \linenumbers

\section{Introduction}\label{intro}
Gamma spectrometry in environments with high background radiation is a challenging task, especially for identification of nuclides emitting gamma rays in the lower energy range of below a few hundred keV and low-level activity. Usually, samples are measured in a shielded environment to prevent the presence of background radiation in the resulting gamma ray spectra. 

In the case of in-situ measurements, shielding also suppresses the low-energy gamma rays resulting from nuclides of interest, not just the background itself. A possible solution is the application of Compton suppression technologies \cite{hpge1} \cite{hpge2} \cite{hpge3} or replacement of the in-situ measurements by sampling techniques \cite{hpge4}. 

This article presents a method for in-situ measurements of gamma ray spectra in the core of a TRIGA reactor focussing on (but not limited to) the nuclide $^{135}$Xe with a main (intensity = 90\%) gamma ray energy of 249.8 keV \cite{lund}.\\\\
The TRIGA Mark-II reactor in Vienna was built by General Atomic in the years 1959 through 1962 and went critical for the first time on March 7, 1962. Since this time the operation of the reactor has averaged at 200 days per year without any longer outages. The reactor core currently consists of 83 fuel element arranged in an annular lattice. Due to the low operating power fuel burn-up occurs slowly and most of the initial loaded core elements are still part of the core \cite{trigapdf}.

The reactor uses a mixed core consisting of three different types of fuel elements (FE): 20\% enriched Aluminium (Al) clad fuel, 20 \% enriched Stainless Steel (SS) clad fuel and 70\% enriched  SS clad FLIP (\emph{Fuel Lifetime Improvement Program}) fuel. The fuel consists of 8 wt \% Uranium, 1 wt \% Hydrogen and 91 wt \% Zirconium with the Zirconium Hydride (ZrH) being the main moderator \cite{trigapdf}.

\begin{figure}[htbp]
\begin{center}
\includegraphics[width=5.5in]{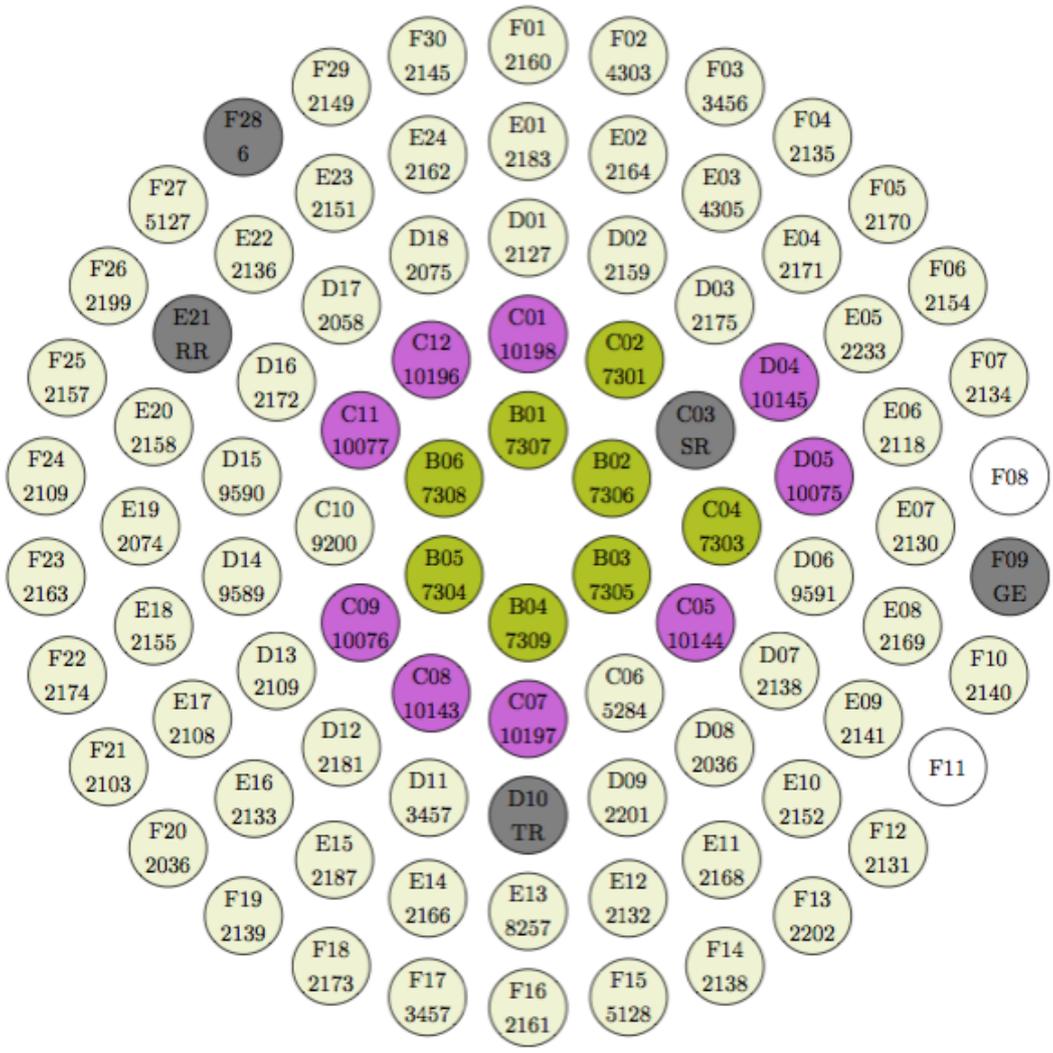}
\caption{TRIGA MArk II Vienna core layout}
\label{core}
\end{center}
\end{figure}

The current layout of the TRIGA core is shown in fig. \ref{core}.
During reactor operation, the generation of fission products results in additional absorption (reactor poisoning). The dominant poisoning nuclide in the TRIGA reactor in Vienna is  $^{135}$Xe. During regular operation, the reactor in Vienna accumulates enough  $^{135}$Xe to make corrections of the regulating rod positions needed to compensate this poisoning. After four days of regular operation, the negative reactivity caused by Xenon poisoning equals about 0.11 cent  \cite{trigapdf}.\newpage
\section{Experimental setup}\label{setup}
\begin{figure}[htbp]
\begin{center}
\includegraphics[width=3.5in]{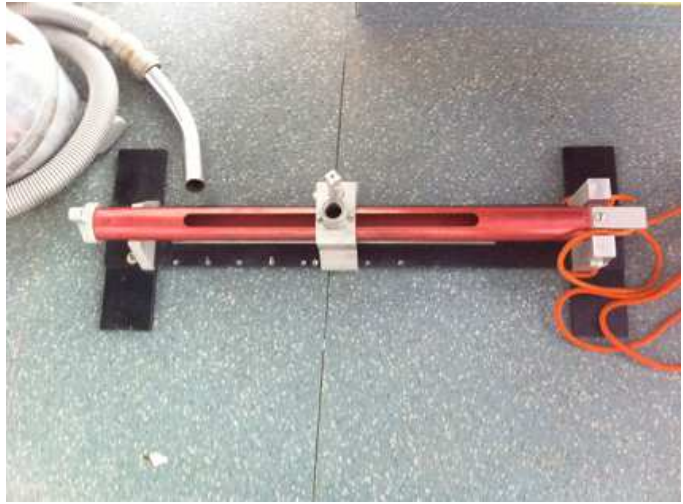}
\caption{Fuel rod scanning device}
\label{figaluelox}
\end{center}
\end{figure}

\begin{figure}[htbp]
\begin{center}
\includegraphics[width=3.5in]{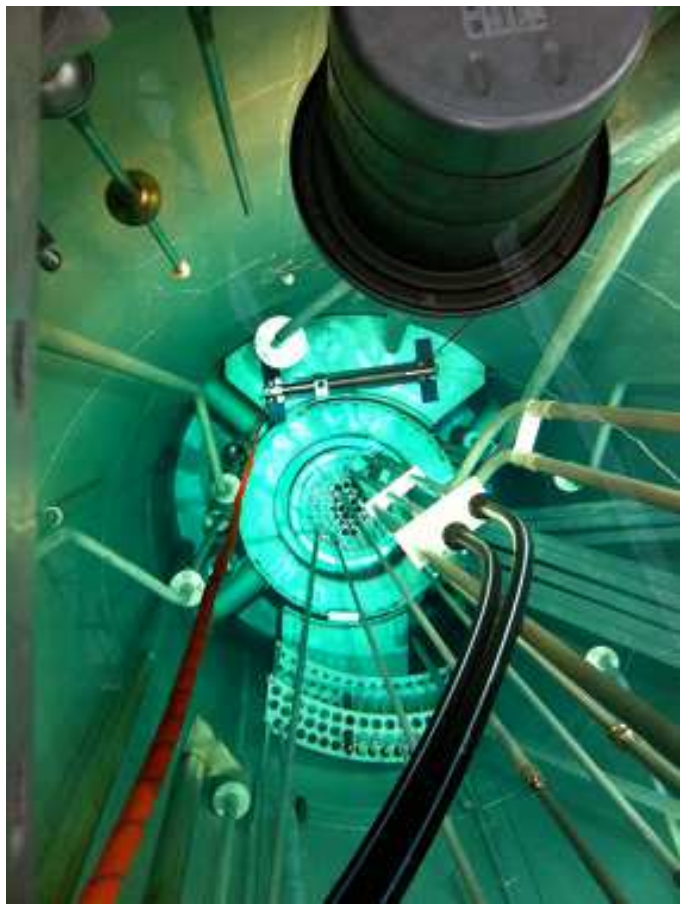}
\caption{Fuel rod scanning device placed onto the thermal column (final position for measurements)}
\label{figtc}
\end{center}
\end{figure}

The gamma spectrometry setup for this experiment consisted of an High Purity Germanium detector (Princeton Gamma Tech) together with a Multi Channel Analyzer manufactured by GBS (MCA166). The system was connected to a personal computer via serial bus for recording purposes.\\\\
For the spectrometrical measurements, one fuel rod has been transferred from its original position in the annular lattice into a fuel rod scanning device made out of Aluminium which is shown in fig. \ref{figaluelox}. The cylinder in the middle holds the fuel element and can be moved from an upright position into a position parallel to the ground plane as shown in fig. \ref{figtc}. The upright position is used for easier fuel rod handling (transfer from its position within the annular lattice onto the thermal column for measurement).

The cut in the upper side of the cylinder holding the fuel rod is equipped with a moveable device including a winding to fix a hollow Aluminium cylinder. This provides an air-filled tube from the bottom of the reactor tank up to the water surface where the detector is placed, thus forming a kind of beamline. This beamline is positioned directly under the detector mounting to ensure a mostly undisturbed beam between the fuel element and the detector.\\\\
The beam is weakened by the bottom of the Aluminium tube (Aluminium winding) and collimated by a lead slab with a 5mm drilling in the center positioned at the bottom of the Aluminium tube (directly above the winding). After placing the fuel rod scanning device onto the thermal column (see fig. \ref{figtc}), the Aluminium tube is screwed into the winding at the top of the device creating a temporal beamline up to the gamma detector mounted at the top. 
\begin{figure}[htbp]
\begin{center}
\includegraphics[width=3.5in]{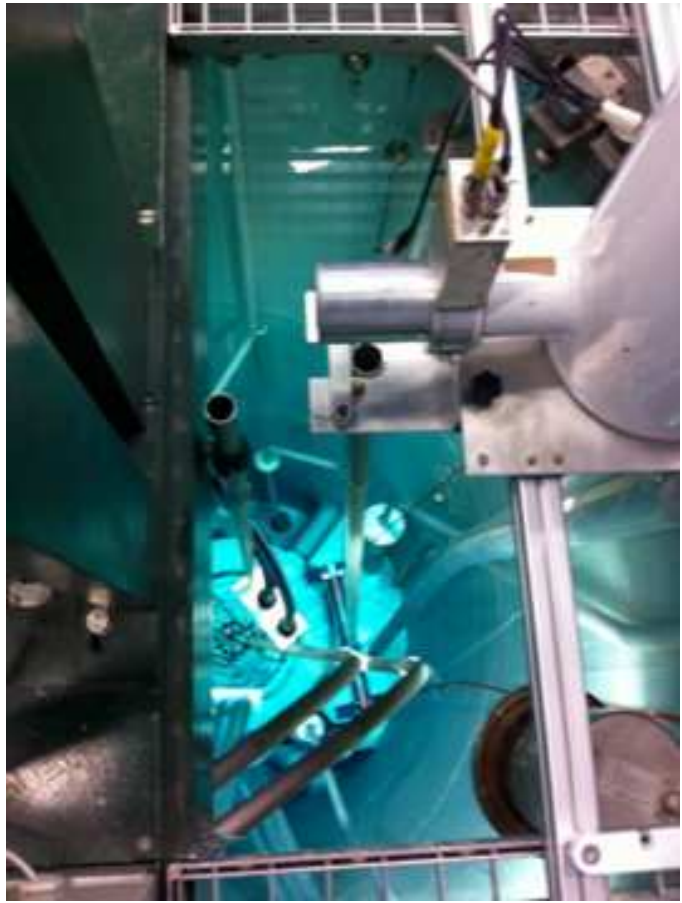}
\caption{HPGe detector over the temporary beamline}
\label{hpge}
\end{center}
\end{figure}
An overview of the upper part of the measurement setup (upper part of the beamline tube and the detector mounting) can be seen in fig. \ref{hpge}.\\\\
Measurements were done with one fuel element withdrawn from the core in February 2001 for testing purposes (element number 2176 from the spent fuel storage within the tank) and with one selected element from the F ring (F5, element number 2170).
\section{Calculation of the time-dependent $^{135}$Xe inventory}\label{xenon}
The thermal absorption cross section of $^{135}$Xe is 2.5 $\times$ 10$^6$ barns. It is produced either directly from fission (with yield $\gamma_{Xe}$) or by beta decay of $^{135}$Te (with yield $\gamma_{Te}$) via beta decay of $^{135}$I (with yield $\gamma_I$).  As the half-life time of 135-Te is just 19s compared to the half-life times of $^{135}$I and $^{135}$Xe (6.6 hours and 9.1 hours, respectively) the production yield of Iodine $\gamma_I$ can be safely set to $\gamma_{Te}$. A comprehensive overview about half life times and fission yields for the nuclides mentioned above are listed in table \ref{tabconstants}.

\begin{table}[htdp]
\caption{Values of constants used for all subsequent calculations}
\begin{center}
\begin{tabular}{|l|l|l|}
\hline
Constant & Desrciption & Value\\
\hline
\hline
$\lambda_{I}$ & $^{135}$I decay constant [s$^{-1}$] & 2.90E-05 \cite{emendoerfer}\\
\hline
$\lambda_{Xe}$ & $^{135}$Xe decay constant [s$^{-1}$]& 2.10E-05 \cite{emendoerfer}\\
\hline
$\sigma_{a,I}$ & absorption cross section for $^{135}$I [cm$^2$]& 2.20E-27 \cite{emendoerfer}\\
\hline
$\sigma_{a,Xe}$ & absorption cross section for $^{135}$Xe [cm$^2$]& 2.50E-19 \cite{emendoerfer}\\
\hline
$\gamma_{Sb}$ & $^{135}$Sb production yield per fission& 1.50E-03 \cite{emendoerfer}\\ 
\hline
$\gamma_{Te}$ & $^{135}$Te production yield per fission & 3.13E-02 \cite{emendoerfer}\\
\hline
$\gamma_I$ &  $^{135}$I production yield per fission & 3.03E-02 \cite{emendoerfer}\\
\hline
$\gamma_{Xe}$ & $^{135}$Xe production yield per fission & 2.40E-03 \cite{emendoerfer}\\
\hline
\end{tabular}
\end{center}
\label{tabconstants}
\end{table}%

The time dependent Iodine and Xenon concentrations $I(t)$ and $Xe(t)$ are given by \cite{stacey}
\begin{equation}\label{xe01}
I(t) = \frac{\gamma_{Te}\Sigma_f \Phi}{\lambda_I}\left( 1-e^{-\lambda_I t} \right) + I(0)e^{-\lambda_I t}
\end{equation}
\begin{align}\label{xe02}
Xe(t) = \frac{\left(\gamma_{Te} + \gamma_{Xe}\right)\Sigma_f \Phi}{\lambda_{Xe}+\sigma_a^{Xe}\Phi}\left[1- e^{-\left( \lambda_{Xe} + \sigma_a^{Xe} \Phi\right)t}\right]\\
+ \frac{\gamma_{Te}\Sigma_f \Phi - \lambda_I I(0)}{\lambda_{Xe}-\lambda_I + \sigma_a^{Xe}\Phi} \left[ e^{-\left(\lambda_{Xe}+\sigma_a^{Xe}\Phi\right)t} - e^{-\lambda_I t} \right] +Xe(0)e^{-\left(\lambda_{Xe} + \sigma_a^{Xe} \Phi\right)t} \nonumber
\end{align}
To provide analytical data for Iodine and Xenon development over time including reactor operation periods equations \ref{xe01} to \ref{xe02} have been used to calculate $I(t)$ and $Xe(t)$. For $I(t)$, the analytical solution has been used. For calculation of $Xe(t)$, equation \ref{xe01} has been integrated numerically.
\section{Operation history in the relevant time period}\label{history}
Using the setup described in section \ref{setup}, gamma ray spectra have been measured from Jan 28, 2011 to Feb 4 2011 starting after reactor shutdown each day. On each measurement day, the reactor operating history has been recorded to be able to reproduce the Iodine and Xenon production and decay history. An overview about the operation history can be found in table \ref{tabxereactorhistory}. On the weekend between Jan 28 and Jan 31 the reactor remained in shutdown state. The measurements were continued over the whole weekend.

\begin{figure}[htbp]
\begin{center}
\includegraphics[width=5.5in]{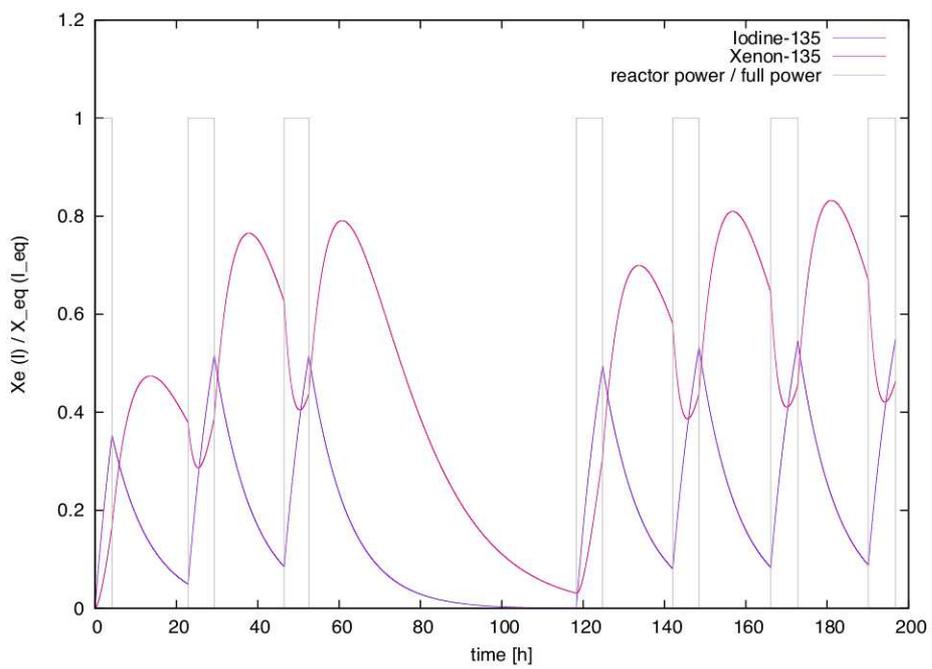}
\caption{Theoretical Iodine and Xenon development with reactor operation history between 2011-01-28 and 2011-02-04}
\label{fig-xe-theory}
\end{center}
\end{figure}

The inventory of $^{135}$Xe within the measurement time period has been calculated as described in section \ref{xenon}. A graphical overview of this inventory together with the reactor operation history in the relevant time period is shown in fig. \ref{fig-xe-theory}.

\begin{table}[htdp]
\caption{Reactor operation between Jan 28, 2011 and Feb 3, 2011}
\begin{center}
\begin{tabular}{|c|l|c|c|c|c|}
\hline
\bf{Date} & \bf{Weekday} & \bf{Startup} & \bf{Shutdown} & \bf{Start} & \bf{End}\\
\hline
2011-01-26 & Wednesday & 10:57 & 15:06 & - & - \\
\hline
2011-01-27 & Thursday & 09:49 & 16:15 & - & - \\
\hline
2011-01-28 & Friday & 09:26 & 15:30 & 16:45 & - \\
\hline
2011-01-29 & Saturday & - & - & - &  - \\
\hline
2011-01-30 & Sunday & - & - & - &  14:38\\
\hline
2011-01-31 & Monday & 09:16 & 15:45 & 16:03 & 08:46 \\
\hline
2011-02-01 & Tuesday & 09:00 & 15:27 & 15:58 & 08:45 \\
\hline
2011-02-02 & Wednesday & 09:05 & 15:45 & 16:48  & 08:00 \\
\hline
2011-02-03 & Thursday & 09:04 & 15:45 & 16:21  & 08:00 \\
\hline
\end{tabular}
\end{center}
\label{tabxereactorhistory}
\end{table}%

\section{The peak search and evaluation algorithm}
\label{peakalgo}
The algorithm is based on the peak finding algorithm of Mariscotti \cite{mariscotti}. This peak finding algorithm is based on the assumption that peaks can be described by Gaussian functions and the background may be approximated by a linear function within short intervals. In such an interval the number of counts as a function of the channel number $x$ is
\begin{equation}
\label{mariscotti1}
N(x) = G(x) + B + Cx
\end{equation}
where $G(x)$ represents a Gaussian function if a peak is present (and zero otherwise), $B$ and $C$ are constants describing the background. If one assumes that $N(x)$ is a continuous function, the second derivative $N''(x)$ becomes independent of the background and vanishes for any interval in which there is no peak \cite{mariscotti}. Therefor peaks are located wherever $N''(x)$ $\ne$ 0.

After generating the second difference $D''(x)$ by using the second-difference approximation
\begin{equation}
D(x) := N''(x) = N_s(x-1) + N_s(i+1) - 2 N_s(i)
\end{equation}
peak candidates are identified by the condition
\begin{equation}
\label{eqcandidate}
|D(x)| > c\quad U(x)
\end{equation}
where $U(x)$ is the uncertainty in the original measurement and $c$ is a user defined parameter (threshhold) which reflects search sensitivity.\\\\
Application of second-derivative methods for peak finding is limited by statistical noise introduced by the measurement equipment. This noise has to be filtered out before peak searching to get reasonable results. A suitable filter for this task is binomial smoothing \cite{marchand}, also called the moving average.

\begin{figure}[htbp]
\begin{center}
\includegraphics[width=5.5in]{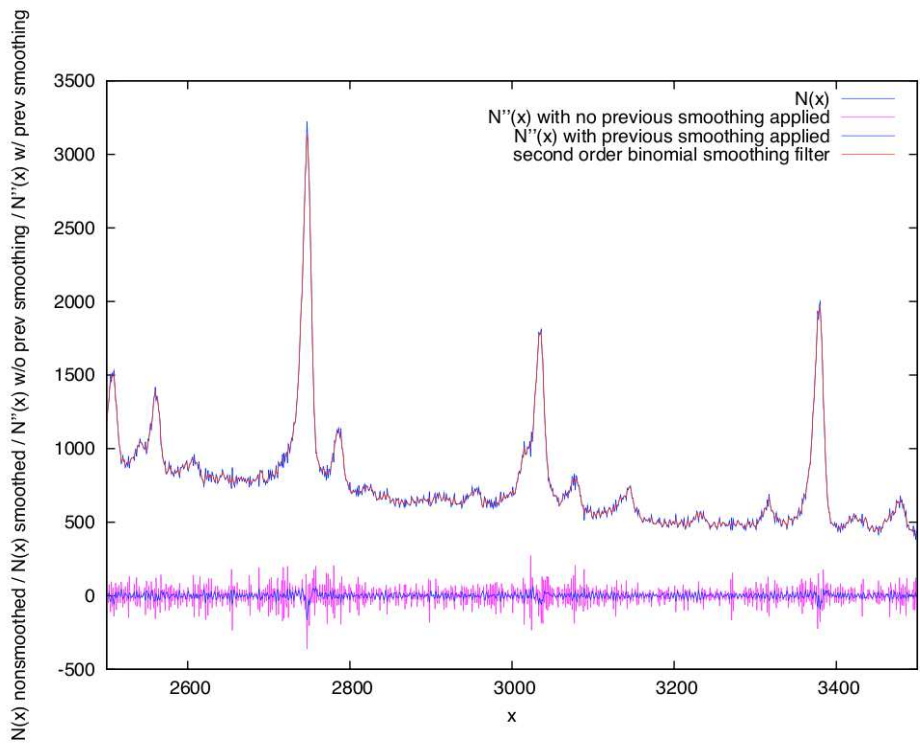}
\caption{N(x) and N''(x) for non-smoothed versus smoothed input data}
\label{figsmoothing}
\end{center}
\end{figure}

\begin{figure}[htbp]
\begin{center}
\includegraphics[width=5.5in]{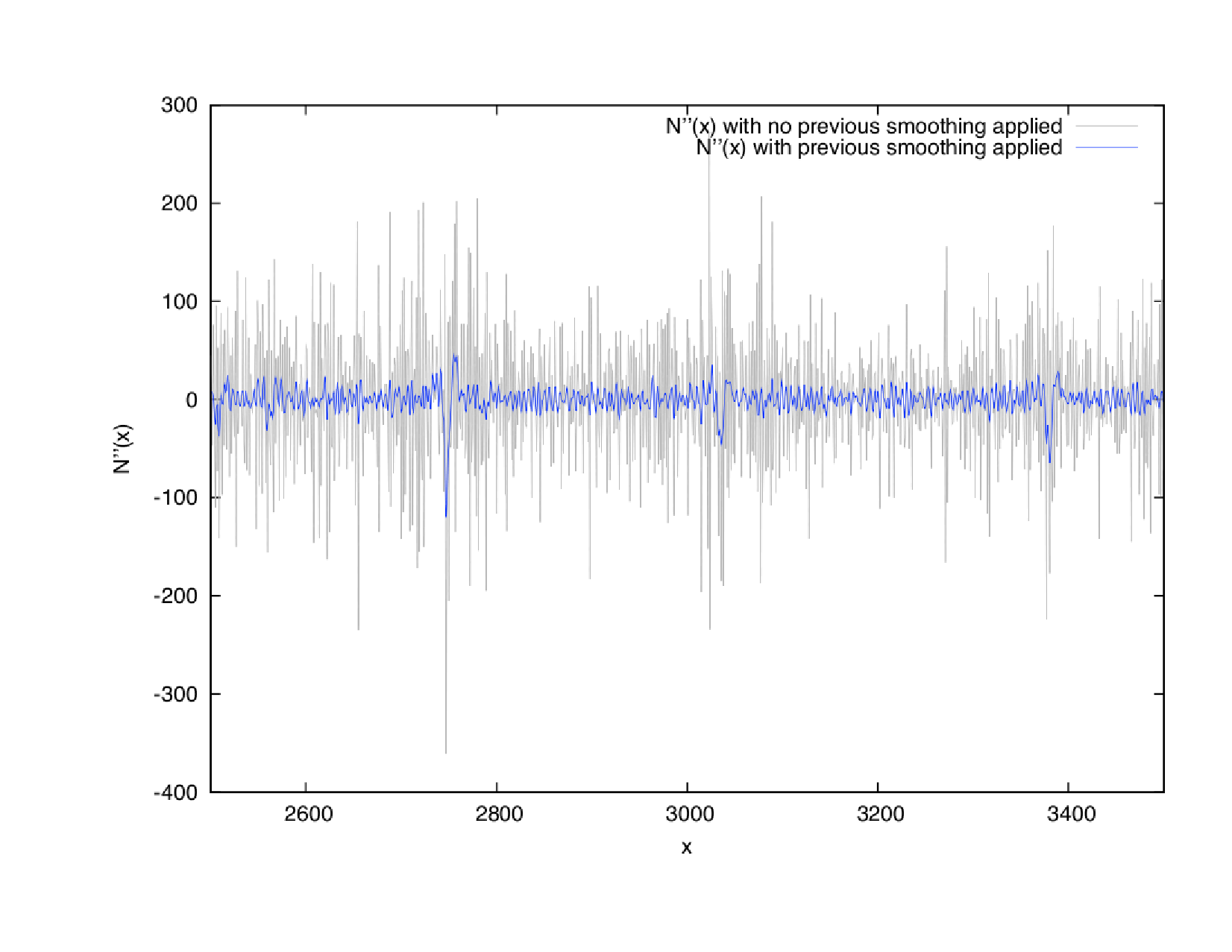}
\caption{Close-up: comparison of the second derivatives of previously smoothed vs. non-smoothed input data, fourth order binomial smoothing filter}
\label{figsmoothing3}
\end{center}
\end{figure}

For example consider fig. \ref{figsmoothing}. In the upper part of the plot the input data $N(x)$ (blue) and the smoothed data $N_s(x)$ (red) are shown. In the lower part the second derivatives $N_s''(x)$ for previously smoothed input data (blue) and the non-smoothed data $N''(x)$ (fuchsia) are shown. It is clearly visible that the second derivative of the previously non-smoothed data is dominated by noise and cannot be analysed easily. Figure \ref{figsmoothing3} shows a close-up of the second derivatives for a fourth order binomial smoothing filter.
\\\\
Choosing the order of the binomial smoothing filter is essential to maintain data integrity during evaluation and analysis. The choice is made based on efficiency of peak detection during a systematic investigation of peaks detection with various input data (high, medium and low dose rate measurements short after reactor shutdown, after about a day and after three days without operation).
\section{Experimental results}\label{results}
The most stable spectra were recorded on Feb. 1, 2011 and Feb. 3, 2011. The number of Xenon and Iodine nuclei in this period have been calculated according to equations (\ref{xe01}) and (\ref{xe02}). The results are shown in fig. \ref{fig-xe-theory}, together with the reactor operation history during this time.
To find the peak regions of interest, a histogram of all peaks according to the algorithm presented in chapter \ref{peakalgo} is generated. Each spike represents one identified peak location in these spectra.

The histogram strongly depends on the given threshold $c$ as defined in chapter \ref{peakalgo}. If the threshold is too small, too much minima are counted especially in the lower energy band resulting in a noisy analysis for low energy peaks, like for one of the nuclides in question ($^{135}$Xe at 249.8 keV). This nuclide is then no longer identified correctly. If defining the threshold too small, some weakly defined peaks like the Iodine peaks at 1131.5 keV and 1260 keV are no longer identified correctly. 

\begin{figure}[htbp]
\begin{center}
\includegraphics[width=5.5in]{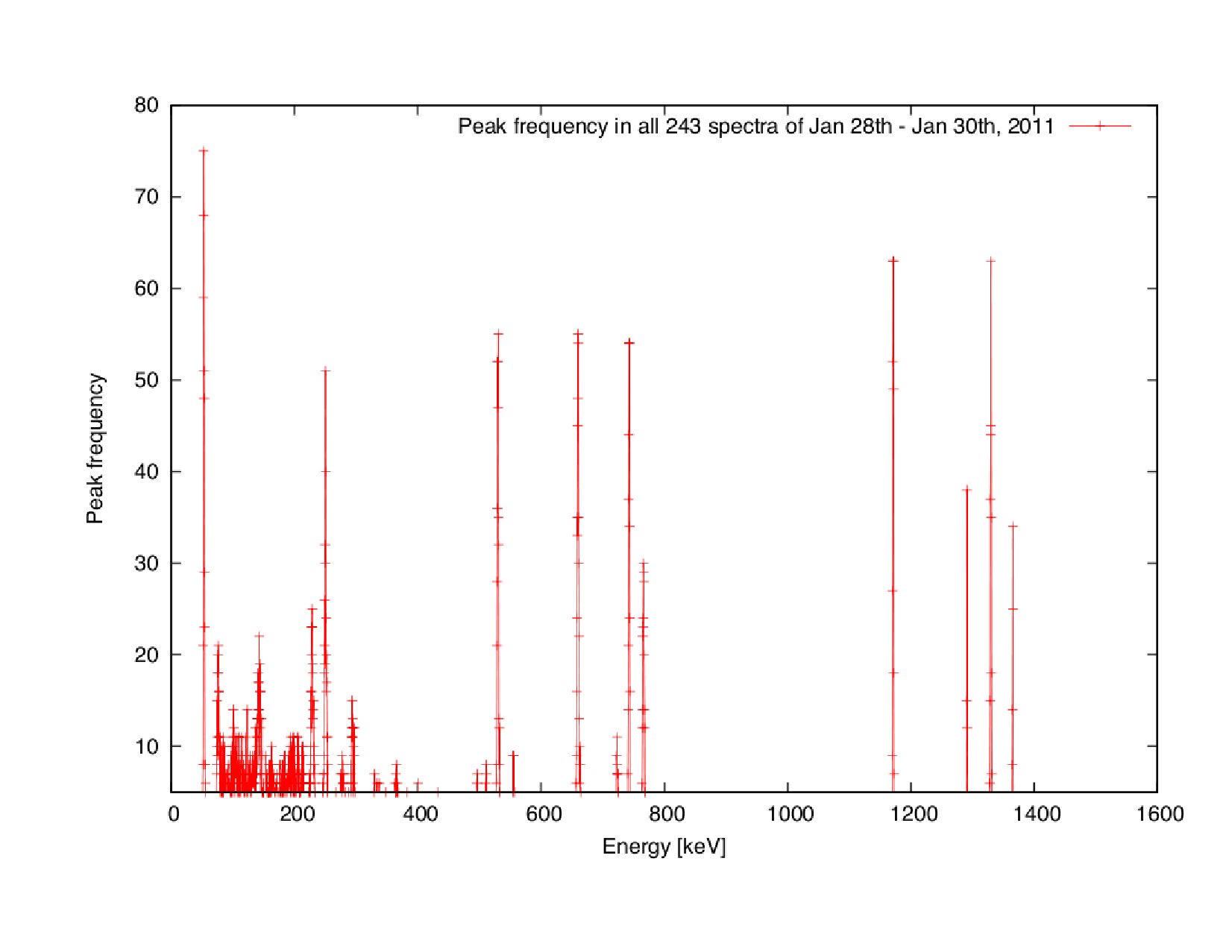}
\caption{Peak frequency in all 243 spectra of Jan. 28th - Jan. 30th, 2011 ($c$ = 0.08$c_{gm}$)}
\label{figfrequencyweekend}
\end{center}
\end{figure}

\begin{figure}[htbp]
\begin{center}
\includegraphics[width=5.5in]{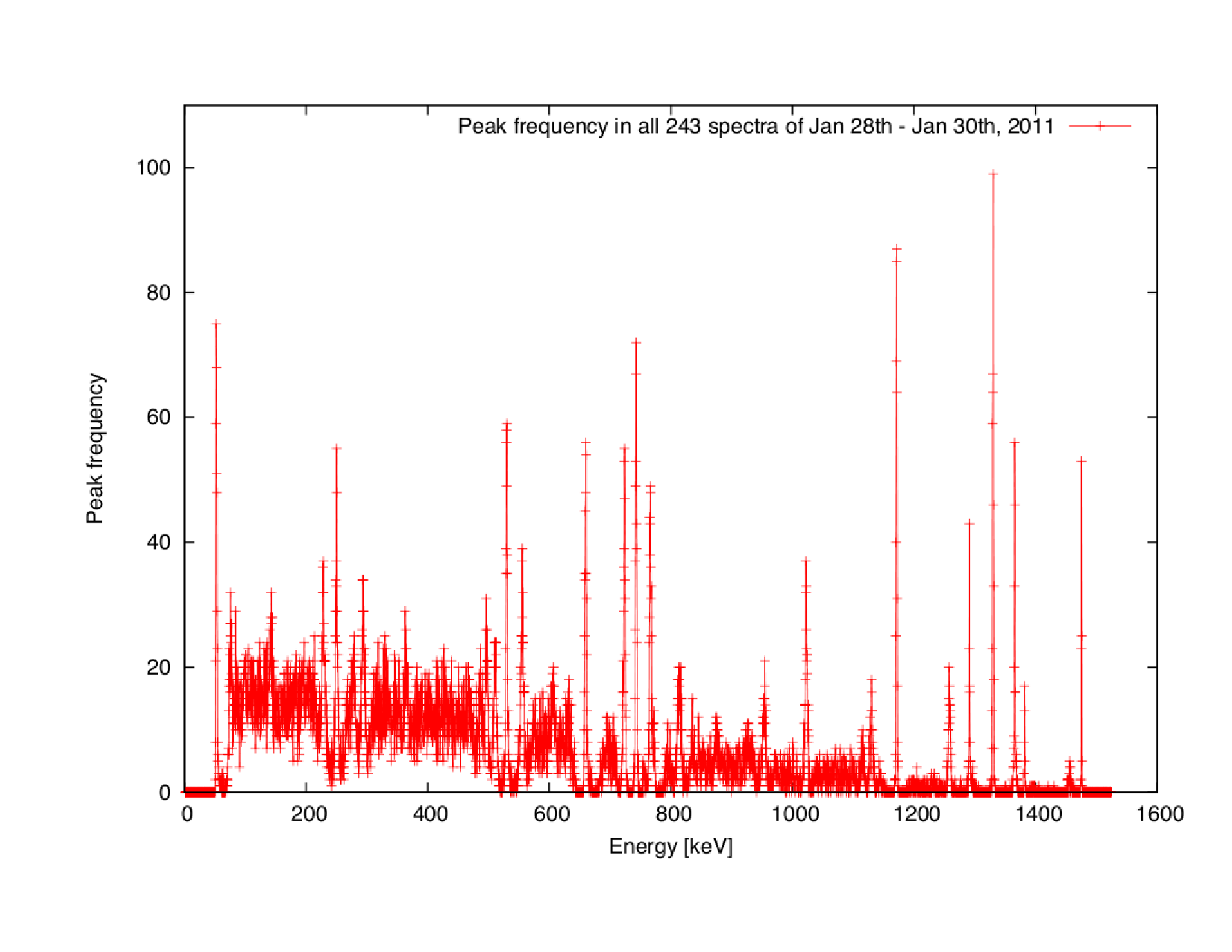}
\caption{Peak frequency in all 243 spectra of Jan. 28th - Jan. 30th, 2011 ($c$ = 0.02$c_{gm}$)}
\label{fig-hist-th2}
\end{center}
\end{figure}

\begin{figure}[htbp]
\begin{center}
\includegraphics[width=5.5in]{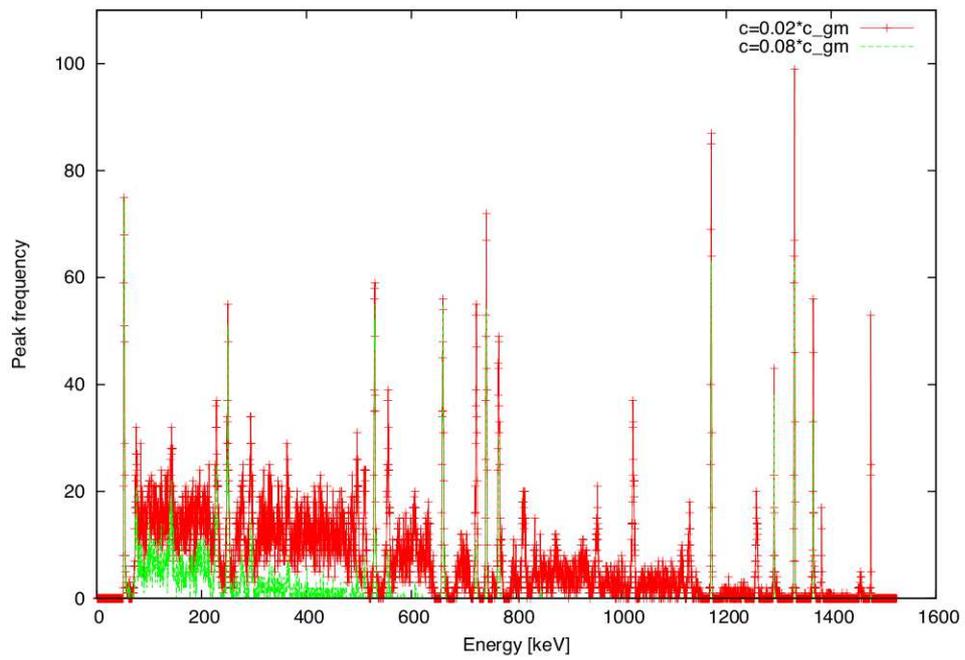}
\caption{Peak frequency in all 243 spectra of Jan. 28th - Jan. 30th, 2011 (combined)}
\label{fig-hist-two}
\end{center}
\end{figure}

For this purposes, two runs of the same algorithms with different sensitivities have been done. The sensitivity is given in units of the global minimum (all peaks, all spectra) $c_{gm}$. For correct identification of the low-energy peaks, a threshold of $c$ = 0.08$c_{gm}$ has been used, for the higher energy peaks a threshold of $c$ = 0.02$c_{gm}$ has been used.
The generated histogram for the 243 spectra in this time range for $c$ = 0.08$c_{gm}$ is shown in fig. \ref{figfrequencyweekend}, for $c$ = 0.02$c_{gm}$ in fig. \ref{fig-hist-th2} and the two combined histograms outlining their differences is shown in fig. \ref{fig-hist-two}.

Fig. \ref{fig-hist-th2} clearly shows the noise in the lower energy range. The amount of peaks identified makes it impossible to clearly distinct between peaks located very closely to each other. On the other hand, it can clearly be seen that many of the peaks in the higher energy range are not correctly identified by using a too small threshold, but for $c$=0.08$c_{gm}$ they are identified correctly.

{\small \begin{table}[htdp]
\caption{Identified peaks in the 243 spectra measured between Jan. 28th - Jan. 30th, 2011 (excerpt)}
\begin{center}
\begin{tabular}{|c|c|c|c|c|}
\hline
peak \# & $x_1$ [keV] & $x_2$ [keV] & $\bar{x}$ [keV] & Nuclide\\
\hline
4 ($c=0.02c_{gm}$) & 137.6 $\pm 0.01$ & 149.0 $\pm 0.01$ & 143.3 $\pm 0.07$& $^{235}$U (143.8 keV)\\
\hline
5 ($c=0.02c_{gm}$) & 180.2 $\pm 0.01$& 188.6 $\pm 0.01$& 184.4 $\pm 0.07$& $^{235}$U (185.7 keV)\\
\hline
9 ($c=0.02c_{gm}$) & 243.7 $\pm 0.01$& 257.1 $\pm 0.01$& 250.4 $\pm 0.07$& $^{135}$Xe (249.7 keV)\\
\hline
31 (both) & 507.9 $\pm 0.01$& 514.3 $\pm 0.01$& 511.1 $\pm 0.07$& annihilation (511 keV)\\
\hline
35 (both) & 654.4 $\pm 0.01$& 667.5 $\pm 0.01$& 660.9 $\pm 0.07$& $^{137}$Cs (661.6 keV))\\
\hline
70 ($c=0.08c_{gm}$) & 1124.8 $\pm 0.01$& 1135.7 $\pm 0.01$& 1130.2 $\pm 0.07$& $^{135}$I (1131 keV))\\
\hline
71 ($c=0.08c_{gm}$)& 1166.2$\pm 0.01$ & 1176.6 $\pm 0.01$& 1171.4 $\pm 0.07$& $^{60}$Co (1171 keV)\\
\hline
73 ($c=0.08c_{gm}$)& 1287.9$\pm 0.01$ & 1295.4 $\pm 0.01$& 1291.6 $\pm 0.07$& $^{41}$Ar (1291 keV)\\
\hline
74 ($c=0.08c_{gm}$)& 1324.7 $\pm 0.01$& 1335.1 $\pm 0.01$& 1329.9 $\pm 0.07$& $^{60}$Co (1332 keV)\\
\hline
75 ($c=0.08c_{gm}$)& 1361.3 $\pm 0.01$& 1370.2 $\pm 0.01$& 1365.8 $\pm 0.07$& $^{24}$Na (1365 keV))\\
\hline
\end{tabular}
\end{center}
\label{default}
\end{table}}\newpage
\section{Results and Discussion}
\begin{figure}[htbp]
\begin{center}
\includegraphics[width=5.5in]{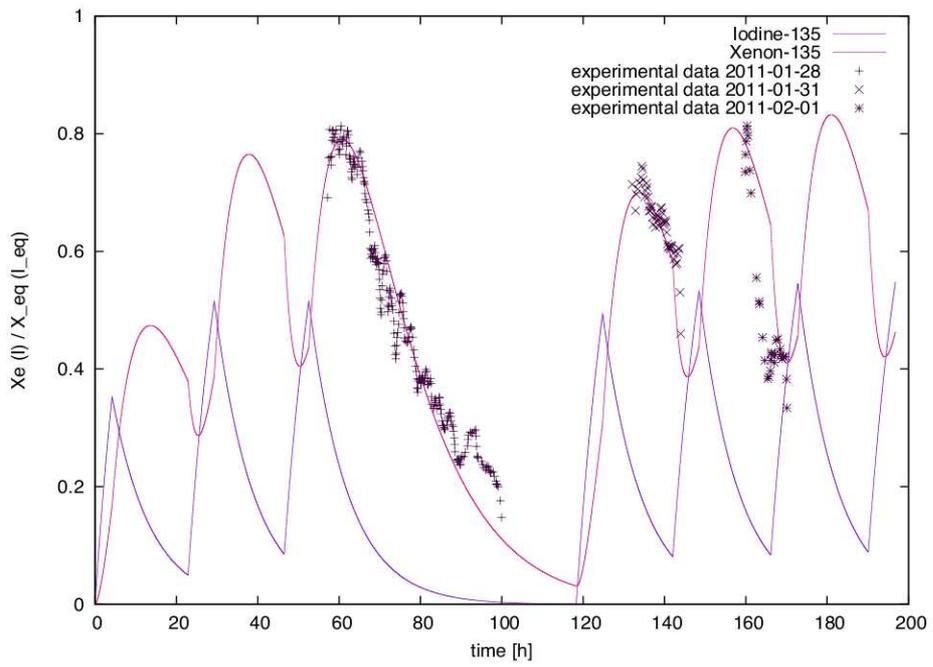}
\caption{Xenon development (numerical results) and experimental data}
\label{figxe-xe-results}
\end{center}
\end{figure}

The algorithm described in chapter \ref{peakalgo} has been applied to all 1277 spectra recorded. Not all of them were usable due to either poor counting statistics and/or high dead times. Especially the latter has been a problem due to the high gamma ray background generated in presence of the reactor and the fuel element mounted in the measurement position. Two measurement batches were good enough to be evaluated: the measurement over the weekend from Jan 28th to Jan 31st and the overnight measurement starting on Jan 31st.\\\\
The results, which are shown graphically in fig. \ref{figxe-xe-results}, show that the $^{135}$Xe inventory can be measured using the data processing techniques described above even in the presence of high-background radiation and therefor in-situ measurement of gamma ray spectra in TRIGA reactors are not only possible, but also relatively easy to obtain.
\newpage


%% References


\bibliographystyle{model1-num-names}
\bibliography{<your-bib-database>}



\end{document}